\documentclass[prd,twocolumn,superscriptaddress,altaffilletter,nofootinbib]{revtex4}

\usepackage{cancel}
\usepackage{graphicx}
\usepackage{amsmath}
\usepackage{amssymb}
\usepackage{graphicx,epsfig}
\newtheorem{theorem}{Theorem}
\newtheorem{corollary}{Corollary}[theorem]

\newcommand{\be}{\begin{equation}}
\newcommand{\ee}{\end{equation}}
\newcommand{\bea}{\begin{eqnarray}}
\newcommand{\eea}{\end{eqnarray}}
\newcommand{\der}{\partial}
\newcommand{\vphi}{\varphi}
\newcommand{\bet}{\begin{theorem}}
\newcommand{\eet}{\end{theorem}}
\newcommand{\bec}{\begin{corollary}}
\newcommand{\eec}{\end{corollary}}

\begin{document}

%%%%%%%%%%%%%%%%%%%%%%%%%%%%%%%%%%%%%%%%%

%%%%%%%%%%%%%%%%%%%%%%%%%%%%%%%%%%%%%%%%%%%%%%%%%%%%%%%%%%%%%%%%%%%%%%%%%%%%%%%%%%%%%%%%%%%%%

\title{On the equivalence between S\'aez-Ballester theory and Einstein-scalar field system}

%%%%%%%%%%%%%%%%%%%%%%%%%%%%%%%%%%%%%%%%%%%%%%%%%%%%%%%%%%%%%%%%%%%%%%%%%%%%%%%%%%%%%%%%%%%%%

%%%%%%%%%%%%%%%%%%%%%%%%%%%%%%%%%%%%%%%%%%%%%%%%%%%%%%%%%%%%%%%%%%%%%%%%%%%%%%%%%%%%%%%%%%%%%%%%%%%%%%%%%%%%%%%%%%%%%%%%%%%%%%%%

\author{Israel Quiros}\email{iquiros@fisica.ugto.mx}\affiliation{Dpto. Ingenier\'ia Civil, Divisi\'on de Ingenier\'ia, Universidad de Guanajuato, Gto., M\'exico.}

\author{Francisco Antonio Horta-Rangel}\email{anthort@ugto.mx}\affiliation{Dpto. Ingenier\'ia Civil, Divisi\'on de Ingenier\'ia, Universidad de Guanajuato, Gto., M\'exico.}

%%%%%%%%%%%%%%%%%%%%%%%%%%%%%%%%%%%%%%%%%%%%%%%%%%%%%%%%%%%%%%%%%%%%%%%%%%%%%%%%%%%%%%%%%%%%%%%%%%%%%%%%%%%%%%%%%%%%%%%%%%%%%%%

%%%%%%%%%%%%%%%%%%%%%%%%%%%%%%%%%%%%%

\begin{abstract}
Here we discuss a topic that comes up more often than expected: A same theory or theoretical model arises in two different presentations which are assumed to be actually different theories so that these are independently developed. Sometimes this leads to an unwanted doubling of the results. In this paper we illustrate this issue with the example of two apparently different gravitational theories: (i) the (minimally coupled) Einstein-massless-scalar system and (ii) the S\'aez-Ballester theory. We demonstrate that the latter is not a scalar-tensor theory of gravity, as widely acknowledged. Moreover, S\'aez-Ballester theory is identified with the Einstein-massless-scalar theory. As illustrations of this identification we show that several known solutions of S\'aez-Ballester theory are also solutions of the Einstein-massless-scalar system and viceversa. Cosmological arguments are also considered. In particular, a dynamical systems-based demonstration of the dynamical equivalence between these theories is given. The study of the asymptotic dynamics of the S\'aez-Ballester based cosmological model shows that there are not equilibrium points which could be associated with accelerated expansion, unless one includes a cosmological constant term or a self-interacting scalar field. This is a well-known result for cosmological models which are based in the Einstein-self-interacting-scalar theory, also known as quintessence.\end{abstract}

%%%%%%%%%%%%%%%%%%%%%%%%%%%%%%%%%%%%

%%%%%%%%%%%%%%%%%%%%%%%%%%%%%%%%%%%%%

\maketitle

%----------------beginning of draft---------------------

\section{Introduction} 

Scalar fields have played a major role in the gravitational theories as well as in the standard model of particles (SMP). Within the framework of the metric theories of gravitation \cite{will-lrr-2014}, the so called scalar-tensor theories of gravity \cite{fujii_book_2004, faraoni-book-2004, clifton-phys-rept-2012, quiros-ijmpd-rev-2019} have received much attention as viable alternatives to general relativity, in the search for solutions to outstanding problems of the latter theory \cite{bamba, okinomou}: dark matter (DM) and dark energy (DE) problems, among others. The close connection of scalar-tensor theories (STT)s and the $f(R)$ modified theories has been investigated as well \cite{faraoni-rev-2010}. 

Scalar-tensor theories of gravity are distinguished by the property that, in addition to the graviton, the scalar field is also a carrier of the gravitational interactions. In general one have to differentiate its use as an additional -- perhaps exotic -- matter field in general relativity, from its use as one of the carriers of the gravitational interactions of matter itself. In this regard, for instance, theories of the kind,\footnote{In this paper we use units where the speed of light and the reduced Planck constant $c=\hbar=1$, while the Newton's constant $G_N=(8\pi)^{-1}$.}

\bea S_{GR}=\frac{1}{2}\int d^4x\sqrt{-g}\left[R-(\der\vphi)^2\right],\label{ems-action}\eea where $\vphi$ is a canonical massless scalar field and we have adopted that $(\der\vphi)^2\equiv g^{\mu\nu}\der_\mu\vphi\der_\nu\vphi$, are not scalar-tensor theories. In this case what we have is GR plus a matter field whose Lagrangian ${\cal L}_\vphi=-(\der\vphi)^2/2$. As a matter of fact the equations of motion (EOM) derived from the action \eqref{ems-action} are the Einstein's equations with a massless scalar field as matter source,

\bea &&G_{\mu\nu}\equiv R_{\mu\nu}-\frac{1}{2}g_{\mu\nu}R=T^{(\vphi)}_{\mu\nu},\nonumber\\
&&\nabla^\lambda T^{(\vphi)}_{\lambda\mu}=0\;\Rightarrow\;\nabla^2\vphi=0,\label{ems-eom}\eea where $\nabla^2\equiv g^{\mu\nu}\nabla_\mu\nabla_\nu$ is the d'Alembert operator and the stress-energy tensor of the scalar field is given by,

\bea T^{(\vphi)}_{\mu\nu}=-\frac{2\delta\left(\sqrt{-g}{\cal L}_\vphi\right)}{\sqrt{-g}\delta g^{\mu\nu}}=\der_\mu\vphi\der_\nu\vphi-\frac{1}{2}g_{\mu\nu}(\der\vphi)^2.\label{sf-set}\eea This is also known as Einstein-massless-scalar (EMS) system\footnote{Here we shall use, interchangeably, the whole name ``minimally coupled EMS'' system and the abbreviated one ``EMS'' system.} which has been studied in detail \cite{bergmann-prd-1957, buchdahl-prd-1959, janis-prl-1968, penney-prd-1968, janis-prd-1969, penney-prd-1969, deser-ann-phys-1970, bekenstein-ann-phys-1974, wyman-prd-1981, agnese-prd-1985, abe-prd-1988, griego-prd-1989, husain-prd-1994, indios-ijmpd-1997, indio-ijmpa-1997, indio-aa-1998, gremm-prd-2000, nandi-mpla-2001, vuille-grg-2007, formiga-prd-2011, tafel-grg-2014, cadoni-prd-2015, cadoni-jhep-2016}.

The EMS system is not a scalar-tensor theory. In contrast, theories given by the following action,

\bea S_{STT}=\int d^4x\sqrt{-g}\left[\phi R-\frac{\omega(\phi)}{\phi}(\der\phi)^2\right],\label{stt-action}\eea where $\omega(\phi)$ is the coupling function, are indeed scalar-tensor theories. The main difference of action \eqref{stt-action} with \eqref{ems-action} is in the non minimal coupling between the scalar field and the curvature, through the term $\phi R$. This coupling entails that the metric and the scalar field both propagate the gravitational interactions (this is why these are called as scalar-tensor theories of gravity in the first place.) The resulting effective gravitational coupling $16\pi G_\text{eff}(\phi)=\phi^{-1}$ is a point dependent quantity. In consequence the scalar field determines the strength of the gravitational interactions point by point. For the gravitational constant measured in Cavendish type experiments one gets \cite{fujii_book_2004, faraoni-book-2004, quiros-ijmpd-rev-2019},

\bea 8\pi G_\text{cav}=\frac{1}{\phi_0}\left[\frac{4+2\omega(\phi_0)}{3+2\omega(\phi_0)}\right],\label{g-newton}\eea where $\phi_0=\phi(t_0)$ is the scalar field evaluated at present cosmic time. Notice that only in the limit $\omega(\phi)\rightarrow\infty$, under the normalization where $\phi_0=1$, the measured gravitational constant coincides with the Newton's constant: $G_\text{cav}\rightarrow G_N=(8\pi)^{-1}$. 

Having in mind these facts, one can easily identify a STT, i. e., one can differentiate these theories from theories where the scalar field is non-gravitational and acts only as a matter source in Einstein's equations \eqref{ems-eom}. But not always STT-s are correctly classified.\footnote{The confusion often comes from the indiscriminate use of the conformal transformation of the metric, $$g_{\mu\nu}\rightarrow\Omega^2g_{\mu\nu},\;g^{\mu\nu}\rightarrow\Omega^{-2}g^{\mu\nu},$$ where the positive function $\Omega$ is called as conformal factor, and its geometrical/physical interpretation. For a detailed discussion of this controversial issue see section 6 of Ref. \cite{quiros-ijmpd-2020}.} This is the case of the so called S\'aez-Ballester theory (SBT) \cite{s-b-theory, saez-prd-1987}. In \cite{quiros-ijmpd-2020} it was demonstrated that SBT is not a scalar-tensor theory but it is just the minimally coupled EMS theory. In spite of this demonstration, several recent works on the subject have been published where this demonstration is ignored and SBT is considered as an STT \cite{sbt-recent-1, sbt-recent-2, sbt-recent-3, sbt-recent-4, indii-2022, sbt-recent-5, sbt-recent-6, sbt-recent-7, sbt-recent-8, sbt-recent-9, sbt-recent-10}. For this reason we feel that further discussion on SBT is required in order to make even clearer the full equivalence between SBT and the EMS system. 

Aim of the present paper will be to set the above discussion on solid mathematical basis. For this purpose, in addition to those arguments that we have already published \cite{quiros-ijmpd-2020}, in this paper we perform a dynamical systems study of the asymptotic cosmological dynamics of SBT-based cosmological models and compare the results with those obtained on the basis of a similar study of EMS-based cosmological models. This is a much more convincing way to show the full equivalence between SBT and EMS system, than just looking for particular solutions in both scenarios and comparing them. Besides, the dynamical systems study of the SBT model, including the case when a self-interaction potential for the scalar field is considered, will establish the limitations of SBT as a model for the dark energy. 

We hope that our study would encourage scientists working in gravitational physics to look for equivalences between seemingly different gravitational theories, so as to evade any doubling of the results.\footnote{We sympathize with the complains in \cite{maria} on the lack of efforts on finding equivalences between seemingly different theories of modified gravity. According to the authors of this bibliographic reference, the lack of efforts in the mentioned direction makes the landscape of related theories larger than what it really is, and makes its classification confusing and misleading.} 

The paper has been organized in the following way. In Sec. \ref{sect-basic}, for completeness of our exposition, we include arguments previously published in \cite{quiros-ijmpd-2020}, which allow identifying SBT with the EMS system. In this section we discuss, as well, on several known solutions of S\'aez-Ballester theory and we show that these are solutions of minimally coupled EMS system and viceversa. In Sec. \ref{sect-cosmo} we discuss on cosmological models that are based in SBT/EMS theories and we show why the SBT-based cosmological models can not explain neither the DM nor the DE problems, as incorrectly claimed \cite{s-b-theory, saez-prd-1987}. In Sec. \ref{sect-ds}, where we include a study of the asymptotic dynamics of these cosmological models through the use of the dynamical systems tools, we show that SBT and EMS asymptotic cosmological dynamics are one and the same. In Sec. \ref{sect-acc-expans} we perform a dynamical systems study of SBT with the inclusion of a cosmological constant. This is the simplest way to get accelerated expansion dynamics in this model, at the cost of losing any motivation for replacement of the much more simple general relativity theory (GRT) by the SBT. Then in Sec. \ref{sect-self-sbt}, also for completeness, we include the investigation of the asymtotic cosmological dynamics of SBT-based cosmological model with a self-interacting scalar field (means a self-interaction potential for the scalar field is included.) Discussion of the results and brief conclusions are given in Sec. \ref{sect-discu}. In this paper we use the following signature of the metric: $(-+++)$.

%------------------------------------------------Basics-------------------------------------------

%%%%%%%%%%%%%%%%%%%%%%%%%%%%%%%%%%%%%%%%%%%%%%%%%%%%%%%%%%%%%%%%%%%%%%%%%%%%%%%%%%%%%%%%%%%%%%%%%%%%%%%%%%%

\section{Identification of S\'aez-Ballester theory and Einstein-massless-scalar system}\label{sect-basic} 

%%%%%%%%%%%%%%%%%%%%%%%%%%%%%%%%%%%%%%%%%%%%%%%%%%%%%%%%%%%%%%%%%%%%%%%%%%%%%%%%%%%%%%%%%%%%%%%%%%%%%%%%%%%

S\'aez-Ballester theory is given by the following action \cite{s-b-theory, saez-prd-1987}:

\bea S_{SBT}=\int d^4x\sqrt{-g}\left[R-\omega\phi^n(\der\phi)^2\right],\label{sbt-action}\eea where $\phi$ is the SBT scalar field while $\omega$ and $n$ are free constant parameters. The SBT EOM that can be derived from the above action read,

\bea &&G_{\mu\nu}=\omega\phi^n\left[\der_\mu\phi\der_\nu\phi-\frac{1}{2}g_{\mu\nu}(\der\phi)^2\right],\nonumber\\
&&2\phi^n\nabla^2\phi+n\phi^{n-1}(\der\phi)^2=0.\label{sbt-eom}\eea According to the authors of the original paper \cite{s-b-theory}, the coupling of the scalar field to the metric would lead to more important departures from GR than the $G_N$-varying theories (strictly speaking the STT-s.) 

That SBT is just GR with a minimally coupled massless scalar field as a source of Einstein's equations -- in simpler words; minimally coupled EMS system -- has been demonstrated in \cite{quiros-ijmpd-rev-2019} and also in \cite{quiros-ijmpd-2020}. Although the demonstration is straightforward, here we include it again since it has been ignored in several papers that have appeared after publication of \cite{quiros-ijmpd-rev-2019, quiros-ijmpd-2020} (see, for instance, Refs. \cite{sbt-recent-1, sbt-recent-2, sbt-recent-3, sbt-recent-4, indii-2022, sbt-recent-5, sbt-recent-6, sbt-recent-7, sbt-recent-8, sbt-recent-9, sbt-recent-10} to quote a few of them.) 

Let us perform the following innocuous redefinition of the SBT scalar field,

\bea \vphi=\frac{2\sqrt\omega}{n+2}\,\phi^\frac{n+2}{2}.\label{sf-redef}\eea After this redefinition, the action \eqref{sbt-action} is transformed into the action \eqref{ems-action}, which corresponds to minimally coupled EMS system. In the same way, under the redefinition \eqref{sf-redef} the SBT EOM \eqref{sbt-eom} transforms into the EMS EOM \eqref{ems-eom}. 

In the bibliography one also encounters works that are based in the so called ``generalized SBT,'' where the scalar field's kinetic term in the action \eqref{sbt-action} is replaced by the more general term \cite{socorro-2010}: 

\bea S^\text{gen}_{SBT}=\int d^4x\sqrt{-g}\left[R-F(\phi)(\der\phi)^2\right],\label{gen-sbt-action}\eea where $F(\phi)$ is an arbitrary function. We should notice that in this case the replacement $\vphi=\int\sqrt{F(\phi)}d\phi$ transforms \eqref{gen-sbt-action} into the EMS action \eqref{ems-action}. This suffices to show that SBT must be identified with EMS theory, contrary to expectation in \cite{s-b-theory}. Based on the latter identification, below we shall look for solutions of the EMS theory on the basis of existing solutions of SBT and viceversa. 

Those who are familiar with the k-essence theories \cite{chiba-2000, mukhanov-prd-2001, chiba-prd-2002, copeland-prd-2003, scherrer-prl-2004, babichev} might think that \eqref{sbt-action} belongs in this class of gravitational theories. The action of k-essence is given by (for simplicity of writing we use the following notation $X\equiv-(\der\phi)^2/2$),

\bea S_{K}=\int d^4x\sqrt{-g}\left[\frac{1}{2}R+K(\phi,X)\right],\label{k-ess-action}\eea where $K(\phi,X)$ is a function of the scalar field and of its kinetic energy density. In the bibliography it is mostly used the following decomposition: $K(\phi,X)=K_1(\phi)K_2(X)$. Although k-essence is not a scalar-tensor theory since the scalar field does not modify neither the gravitational coupling nor the measured value of the gravitational constant, it may have cosmological implications differing from those of Einstein-massless-scalar theory, since perturbations of the k-essence field propagate at a sound speed squared $c^2_s$ different from the one obtained in the EMS system ($c^2_s=1$) \cite{babichev}. 

As discussed in \cite{mukhanov-prd-2001}, the linear case $K_2(X)=aX+b$ in \eqref{k-ess-action}, where $a$ and $b$ are free constants, corresponds to GR with a minimally coupled self-interacting scalar field. Only for non-linear functions $K_2(X)$ can we speak of a k-essence field. The SBT action \eqref{sbt-action} corresponds, precisely, to the linear case $K_2(X)=X$ ($K_1(\phi)=\omega\phi^n$,) so that it is GR with a minimally coupled (massless) scalar field, also known as EMS system.

Let us mention a famous case where ignorance of such a simple scalar field redefinition as in Eq. \eqref{sf-redef}, led to unphysical (incorrect) bounds on the parameters of the theory, which led, in turn, to erroneous physical estimates. In \cite{caroll-prd-2003}, in order to explain the present stage of accelerated expansion of the Universe, a self-interacting Einstein-scalar model was proposed where the kinetic energy density of the scalar field entered with the wrong sign. It was dubbed phantom DE. The obvious problem with this model is the instability due to the negative sing of the energy density, which can be avoided only if the instability time scale is greater than the age of the Universe. In \cite{caroll-prd-2003} arguments were given in support of this possibility. One of the analyzed possibilities yielding appropriate constraints was based on an interaction (effective) Lagrangian ${\cal L}_\text{eff}\propto\phi(\der\phi)^2$. As shown in \cite{cline-prd-2004}, this interaction is an artifact of a noncanonical normalization. It is removed if in ${\cal L}_\text{eff}$ make the innocuous replacement $\phi\rightarrow 4\phi^{3/2}/9$ $\Rightarrow$ $\phi(\der\phi)^2\rightarrow(\der\phi)^2$.

%--------------local sols------------

%==========================================

\subsection{Local solutions}
 
%==========================================

Local spherically symmetric solutions of the EMS system have been found \cite{buchdahl-prd-1959, janis-prl-1968, wyman-prd-1981, agnese-prd-1985}. All of these solutions are really the same but expressed in terms of different coordinates. This has been demonstrated in \cite{indio-ijmpa-1997} for the solutions found in Refs. \cite{janis-prl-1968}, \cite{wyman-prd-1981} and in \cite{nandi-mpla-2001} for the solutions \cite{buchdahl-prd-1959}, \cite{janis-prl-1968}. In \cite{agnese-prd-1985}, in particular, the static, spherically symmetric solution to \eqref{ems-eom} is found to be,

\bea &&ds^2=-\left(1-\frac{2\eta}{r}\right)^\frac{m}{\eta}dt^2+\left(1-\frac{2\eta}{r}\right)^{-\frac{m}{\eta}}dr^2\nonumber\\
&&\;\;\;\;\;\;\;\;\;\;\;+\left(1-\frac{2\eta}{r}\right)^{1-\frac{m}{\eta}}r^2d\Omega^2,\nonumber\\
&&\vphi(r)=\frac{\sigma}{\sqrt{2}\eta}\ln\left(1-\frac{2\eta}{r}\right),\label{agnese-sol}\eea where we use spherical coordinates, $(t,r,\theta,\vphi)$, $d\Omega^2\equiv d\theta^2+\sin^2\theta d\vphi^2$, $\eta=\sqrt{m^2+\sigma^2}$ ($m$ and $\sigma$ are free constants) and 

\bea R=\left(1-\frac{2\eta}{r}\right)^\frac{\eta-m}{2\eta}\,r,\nonumber\eea is the standard radial coordinate. In this case the event horizon at $r=2\eta$ shrinks to a point, thus preventing the formation of a black hole \cite{agnese-prd-1985}.

If we substitute the static, spherically symmetric metric,

\bea ds^2=-e^{\gamma(r)}dt^2+e^{-\gamma(r)}dr^2+e^{\beta(r)}r^2d\Omega^2,\label{stat-spher-line}\eea into the SBT EOM \eqref{sbt-eom}, one gets the same solution for the line element than in \eqref{agnese-sol}, while the SBT scalar field is given by,

\bea \phi(r)=\left[\frac{(n+2)\sigma}{2\sqrt{2\omega}\eta}\right]^\frac{2}{n+2}\left[\ln\left(1-\frac{2\eta}{r}\right)\right]^\frac{2}{n+2}.\label{sbt-sc-f-sol}\eea This can be found as well by directly substituting \eqref{sf-redef} into \eqref{agnese-sol}. Notice that the new free parameters $\omega$ and $n$ play no role in the solution for the line-element. Hence, the physical (also geometrical) results are just the same as in \cite{agnese-prd-1985}.

%-------------------------non static------------------------

\subsubsection{Non-static spherically symmetric solutions}

%-----------------------------------------------------------

The non-static, spherically symmetric solution of the EMS system \eqref{ems-eom} was investigated in \cite{husain-prd-1994}. The solution is given by,

\bea &&ds^2=(at+b)\left[-f^2(r)dt^2+f^{-2}(r)dr^2\right]+R^2d\Omega^2,\nonumber\\
&&\vphi(t,r)=\pm2\sqrt\pi\ln\left[d\left(at+b\right)^{\sqrt{3}}\left(1-\frac{2c}{r}\right)^\frac{\alpha}{\sqrt{3}}\right],\label{husain-sol}\eea where $a$, $b$, $c$ and $d$ are free constants, $\alpha=\pm\sqrt{3}/2,$ and

\bea &&f^2(r)=\left(1-\frac{2c}{r}\right)^\alpha,\nonumber\\
&&R^2=R^2(t,r)=(at+b)\left(1-\frac{2c}{r}\right)^{1-\alpha}r^2.\nonumber\eea 

Although this solution does not shed light on the scalar field collapse problem in asymptotically flat space (the solution is not asymptotically flat), it provides an example of spacetimes with evolving apparent horizons \cite{husain-prd-1994}. We can perform the redefinition \eqref{sf-redef} to find the corresponding solution of the SBT system \eqref{sbt-eom}, but this is a futile intent since, as we have demonstrated, the SBT is one and the same as the EMS system.

%-------------------------------------

\subsubsection{Other local solutions}

%-------------------------------------

There are found in the bibliography wormhole solutions of EMS as well \cite{wh-chinos-2022}. In the latter bibliographic reference Bronnikov-type wormhole is investigated. This wormhole solution is possible thanks to a small departure from standard EMS system: since wormhole requires of exotic matter to form, if in \eqref{ems-action} we replace the sign of the kinetic term ``$-\rightarrow+$,'' the scalar field is phantom-like thus providing the exotic matter required by the wormhole. The wormhole solution is given by \cite{wh-chinos-2022},

\bea &&ds^2=-h(r)dt^2+h^{-1}(r)dr^2+R^2(r)d\Omega^2,\nonumber\\
&&\vphi(r)=\frac{\sqrt{2}q}{\sqrt{q^2-M^2}}\ln f(r),\label{wh-sol}\eea where

\bea &&h(r)=f^{-\frac{2M}{\sqrt{q^2-M^2}}}(r),\nonumber\\
&&R^2(r)=\left(r^2+q^2-M^2\right)f^{\frac{2M}{\sqrt{q^2-M^2}}}(r),\nonumber\\
&&f(r)=\exp\left[\arctan\left(\frac{r}{\sqrt{q^2-M^2}}\right)\right].\nonumber\eea In the above equations $q$ and $M$ are integration constants. When $M=0$, the above solution corresponds to the Ellis wormhole. The wormhole \eqref{wh-sol} connects two asymptotic Minkowski spacetimes with different values of the speed of light, so that the wormhole connects two different worlds. 

Through using the redefinition \eqref{sf-redef}, one can bring the above wormhole solution of EMS system into the corresponding wormhole solution of SBT theory. As a matter of fact we can do that with any solution of EMS theory and also one can bring back any solution of S\'aez-Ballester theory into the corresponding solution of EMS theory. Hence, with the help of the innocuous scalar field redefinition \eqref{sf-redef} one can construct a ``dictionary'' of solutions of either SBT or EMS. This, however, will be a futile exercise since, as already shown, both are one and the same theory. 

There are many other known solutions of SBT theory, for instance Bianchi type solutions \cite{bianchi-1, bianchi-2, bianchi-3, bianchi-4, bianchi-5}, as well as of EMS theory, such as Petrov type \cite{petrov} and rotating solutions \cite{rotat-sol}, etc. So that one may ``translate'' these solutions to the EMS system and to the SBT theory, respectively, without difficulty.

%---------------FRW cosmology--------------

%%%%%%%%%%%%%%%%%%%%%%%%%%%%%%%%%%%%%%%%%%%%

\section{FRW cosmology}\label{sect-cosmo}
 
%%%%%%%%%%%%%%%%%%%%%%%%%%%%%%%%%%%%%%%%%%%%

One of the main physical implications of SBT was to (seemingly) take account of the missing-matter problem \cite{s-b-theory}, presently known as the DM problem. Today we know that S\'aez-Ballester theory can not explain neither the DM nor the DE problems \cite{riess-1998, perlmutter-1999, copeland-rev-2006}. 

In order to show why this theory can not explain these problems, let us write the EOM in terms of the Friedmann-Robertson-Walker (FRW) metric with flat spatial sections,

\bea ds^2=-dt^2+a^2(t)\delta_{ij}dx^idx^j,\label{frw-metric}\eea where $a(t)$ is the dimensionless scale factor and $t$ is the cosmic time. In place of the SBT EOM \eqref{sbt-eom} we shall write the simpler and completely equivalent EMS EOM \eqref{ems-eom},

\bea &&3H^2=\frac{1}{2}\dot\vphi^2,\label{fried-eq}\\
&&2\dot H=-\dot\vphi^2,\label{raycha-eq}\\
&&\ddot\vphi+3H\dot\vphi=0,\label{kg-eq}\eea where $H\equiv\dot a/a$ is the Hubble parameter, the dot accounts for derivative with respect to the cosmic time and only two of the above equations are independent of each other. Straightforward integration of equation \eqref{kg-eq} yields,\footnote{Due to absence of a potential (self-interacting) term, the scalar field $\vphi$ behaves as stiff matter fluid.}

\bea \dot\vphi=\frac{\sqrt{2}k^2}{a^3}\;\Rightarrow\;\rho_K=\frac{\dot\vphi^2}{2}=\frac{k^4}{a^6},\label{vphi-edensity}\eea where $k^2$ is an integration constant. Hence, the kinetic energy density of the scalar field $\rho_K\propto a^{-6}$ dies off much faster than the radiation $\rho_r\propto a^{-4}$ and, obviously, much faster than CDM energy density $\rho_m\propto a^{-3}$. Hence, $\vphi$ may have played a role at early times through replacing the matter bigbang by a stronger stiff-matter dominated bigbang, but not at late time (this includes our present stage of the cosmic expansion.) 

Let us, for completeness, to present a general cosmological solution of EMS/SBT system in the presence of a matter component of the cosmic fluid characterized by energy density $\rho_m$ and barotropic pressure $p_m=(\gamma-1)\rho_m$, where $\gamma$ ($0\leq\gamma\leq 2$) is the barotropic index of the fluid (here we consider its most common values: $\gamma=1$ for dust and $\gamma=4/3$ for radiation). The latter cosmological parameter is related with the equation of state (EOS) parameter $w$ of the fluid: $\gamma=w+1$. In this case the EMS EOM read,

\bea &&3H^2=\rho_m+\frac{1}{2}\dot\vphi^2,\label{fried-eq-ems}\\
&&2\dot H=-\gamma\rho_m-\dot\vphi^2,\label{raycha-eq-ems}\\  
&&\ddot\vphi+3H\dot\vphi=0,\label{kg-eq-ems}\\
&&\dot\rho_m+3H(\rho_m+p_m)=0,\label{cont-eq}\eea where the matter fluid continuity equation \eqref{cont-eq} has been included. Integration of this last equations leads to $\rho_m=M^4a^{-3\gamma}$, where $M^4$ is an integrations constant. Substituting the expressions for $\rho_m$ and $\rho_K$ back into \eqref{fried-eq-ems} one gets the Friedmann equation in the following form:

\bea 3H^2=M^4a^{-3\gamma}+k^4a^{-6}.\label{fried-eq''}\eea If one replaces the cosmic time in this equation by the new variable $v$,

\bea t=\int a^3dv,\label{new-var}\eea one can integrate \eqref{fried-eq''} in quadratures to obtain that,

\bea a(v)=a_0\sinh^{\frac{1}{3(\gamma-2)}}\left[\frac{3(\gamma-2)k^2}{2\sqrt{3}}(v-v_0)\right],\label{cosmo-sol}\eea where $v_0$ is an integration constant and,

\bea a_0\equiv\left[\frac{M^4}{k^4}\right]^{\frac{1}{3(\gamma-2)}}.\nonumber\eea 

The scale factor in \eqref{cosmo-sol} can be written in terms of the cosmic time. Actually, by substituting $a(v)$ from \eqref{cosmo-sol} back into \eqref{new-var} and performing the integration, one gets $t=t(v)$. Then one finds the inverse $v=v(t)$ and substitutes in \eqref{cosmo-sol}. The latter is the general solution of \eqref{fried-eq''}.

%%%%%%%%%%%%%%%%%%%%%%%%%%%%%%%%%%%%%%%%%%%%%%%%%%%%%%%%%%%%%%%%%

\section{Dynamical systems-based demonstration}\label{sect-ds}       

%%%%%%%%%%%%%%%%%%%%%%%%%%%%%%%%%%%%%%%%%%%%%%%%%%%%%%%%%%%%%%%%%

In spite of the apparent simplicity of the EOMs \eqref{fried-eq-ems}--\eqref{cont-eq} (same for the system \eqref{fried-sbt}--\eqref{kg-sbt} below,) as with any system of non-linear second-order differential equations, it is a very difficult task to find exact solutions. Even when an analytic solution can be found it will not be unique but just one in a large set of them \cite{faraoni}. In this case the dynamical systems tools are very useful. These very simple tools give us the possibility to correlate such important concepts like past and future attractors (also saddle equilibrium points) in the phase space, with generic solutions of the cosmological EOMs without the need to analytically solve them.\footnote{For a compact pedagogical introduction to the application of the dynamical system tools in cosmology, specifically in scalar field models, see Ref. \cite{quiros-ejp-2015}.} 

The basic idea of the application of the dynamical systems in the cosmological framework is to map the original set of differential equations which constitute the cosmological EOMs, onto an equivalent phase space spanned by appropriate phase variables. Then the original cosmological EOMs are traded by an autonomous system of ordinary differential equations (ASODE) on the phase space variables. What matters are the critical points (also stationary or equilibrium points) of the obtained ASODE and the study of their stability properties. These (isolated or spiral) critical points can be either source (past attractors) or sink (future attractors) or saddle equilibrium points. One then goes back to the original equations and finds which generic solutions of the cosmological EOMs these critical points in the phase space correspond to. These solutions will describe the asymptotic dynamics of the given cosmological model. 

The idea is to demonstrate that the SBT and the EMS system share the same phase space properties, so that their asymptotic dynamics is the same. This will be a much more solid demonstration of the dynamical equivalence of both theories than just looking for particular classes of solutions, since the critical points in the phase space correspond to those classes of solutions which, in a sense, are ``preferred'' by the cosmological EOMs, i. e., those which decide the past and future (also the intermediate) asymptotic behavior. In this section we perform a dynamical systems study of the EMS system and of the SBT theory, in order to show that the asymptotic cosmological dynamics is the same in both cosmological models. In order to get the details of the procedure we follow in the next sections while performing the dynamical systems analysis, we recommend the reader our pedagogically written paper \cite{quiros-ejp-2015}.

%===================================================

\subsection{Asymptotic dynamics of the EMS system}

%===================================================

Let us start by looking for adequate phase space variables in the case of the EMS system, which is described by the EOMs \eqref{fried-eq-ems}--\eqref{cont-eq}. In this case it is well-known that an adequate bounded variable could be \cite{wands-prd-1998},

\bea x\equiv\frac{\dot\vphi}{\sqrt{6}H},\;-1\leq x\leq 1,\label{x-var}\eea where, since we are focused in expanding Universe exclusively, then we shall consider nonegative $H\geq 0$. This means that negative $x$-s amount to decaying scalar field $\dot\vphi<0$, while positive $x$-s entail growing $\vphi$-s: $\dot\vphi>0$.

In terms of the $x$ variable the cosmological EOMs can be written in the following way. The Friedmann equation \eqref{fried-eq-ems} amounts to the following constraint:

\bea \Omega_m=1-x^2,\label{fried-const}\eea where $\Omega_m\equiv\rho_m/3H^2$ is the normalized (dimensionless) matter energy density. The cosmological EOM \eqref{raycha-eq-ems} is written as,

\bea 2\frac{\dot H}{H^2}=-3\gamma\Omega_m-6x^2,\nonumber\eea which, if take into account the Friedmann constraint \eqref{fried-const}, can be written either as,

\bea 2\frac{\dot H}{H^2}=-3\gamma-3(2-\gamma)x^2,\label{raycha-ems-1}\eea or as,

\bea 2\frac{\dot H}{H^2}=-6+3(2-\gamma)\Omega_m.\label{raycha-ems-2}\eea Meanwhile, the scalar field's EOM \eqref{kg-eq-ems} is written as,

\bea \frac{\ddot\vphi}{H^2}=-3\sqrt{6}x,\label{kg-ems}\eea and the continuity equation \eqref{cont-eq} can be written in the following way:

\bea \frac{\dot\rho_m}{3H^3}=-3\gamma\Omega_m,\nonumber\eea or, equivalently

\bea \dot\Omega_m=-H\Omega_m\left(3\gamma+2\frac{\dot H}{H^2}\right).\label{cont}\eea 

It is clear that, thanks to the Friedmann constraint \eqref{fried-const} only one phase space variable is required: either $x$ or $\Omega_m$. Let us choose first the variable $x$. Hence a phase line is enough to represent the asymptotic cosmological dynamics. The autonomous ordinary differential equation (ODE) reads,

\bea \dot x=-\frac{3}{2}\left(2-\gamma\right)Hx(1-x^2),\nonumber\eea or

\bea x'=-\frac{3}{2}\left(2-\gamma\right)x(1-x^2),\label{x-eq}\eea where the prime denotes derivative with respect to the time variable $\tau=\int Hdt$. If instead of the variable $x$ one wants to work with the normalized energy density, one should consider equations \eqref{cont} and \eqref{raycha-ems-2}. We get that,

\bea \Omega'_m=3(2-\gamma)\Omega_m\left(1-\Omega_m\right).\label{om-eq}\eea One of these equations is enough to uncover the asymptotic dynamics of the EMS system. These are not independent of each other since through the Friedmann constraint \eqref{fried-const} one can transform one equation into the other one.

The critical points of the dynamical system \eqref{om-eq} are: (i) the scalar field's kinetic energy dominated solution (also known as stiff-matter solution), $\Omega_m=0$ $\Rightarrow$ $x^2=1$ $\Rightarrow$ $3H^2=\dot\vphi^2/2$, and (ii) the matter-dominated solution $\Omega_m=1$ $\Rightarrow$ $3H^2=\rho_m$. In order to check the linear stability of these critical points one expands the dynamical system around each one of the critical points and considers only the linear approximation. For the stiff-matter solution $\Omega_m=0$ one substitutes $\Omega_m\rightarrow 0\pm\delta$ in \eqref{om-eq} ($\delta$ is a small perturbation) and keeps only terms $\propto\delta$: $\delta'=3(2-\gamma)\delta$. Straightforward integration of this equation yields $\delta(\tau)=\delta_0\exp[3(2-\gamma)\tau]$ ($\delta_0$ is an integration constant and for standard matter $\gamma\leq 2$), so that the scalar field dominated solution is unstable: the linear perturbation $\delta$ around $\Omega_m=0$ grows up in $\tau$-time. This means that the stiff-matter solution is the past attractor in the phase line. Notice that, since $\Omega_m=0$ $\Rightarrow$ $x^2=1$ $\Rightarrow x=\pm 1$, then in the $x$-phase line there are two stiff-matter past attractors: the one for growing values of the scalar field ($x=-1$) and the other one for decreasing values of $\vphi$ ($x=1$). In a similar way, for the matter-dominated solution $\Omega_m=1$, in Eq. \eqref{om-eq} one substitutes $\Omega_m\rightarrow 1\pm\delta$ and considers the linear approximation. We obtain that $\delta'=-3(2-\gamma)\delta$, whose integration leads to: $\delta(\tau)=\delta_0\exp[-3(2-\gamma)\tau]$. This means that the perturbation decreases with $\tau$-time so that this point is stable. I. e., the matter-dominated solution is the future attractor in the phase line.

Summarizing: In the FRW-EMS scenario the ultra relativistic stiff matter solution is the initial stage of the cosmic history while the matter dominated universe is the final state, so that no accelerated expansion stage can be obtained.

%===========================================================

\subsection{Dynamical equivalence of EMS system and SBT}

%===========================================================

In order to demonstrate that SBT and EMS are dynamically equivalent models, let us write the FRW EOM for the SBT theory:

\bea &&3H^2=\rho_m+\frac{\omega}{2}\phi^n\dot\phi^2,\label{fried-sbt}\\
&&2\dot H=-\gamma\rho_m-\omega\phi^n\dot\phi^2,\label{raycha-sbt}\\
&&\ddot\phi+3H\dot\phi+\frac{n}{2}\frac{\dot\phi^2}{\phi}=0,\label{kg-sbt}\eea where the continuity equation \eqref{cont-eq} is the same in both EMS and SBT models. Following the same procedure as above, let us write a variable of the phase space:

\bea u\equiv\frac{\dot\phi}{\sqrt{6}H}=\frac{\phi'}{\sqrt{6}}.\label{u-var}\eea The Friedmann equation \eqref{fried-sbt} can be written in the following way:

\bea \Omega_m=1-\omega\phi^nu^2.\label{fried-sbt'}\eea From this equation it seems that another phase space variable is required. However, instead of trying to find such a variable adequate to this problem, we go a different way. Let us write the SBT EOMs in terms of the variable $u$. We get,

\bea &&\frac{\dot H}{H^2}=-\frac{3\gamma}{2}\Omega_m-3\omega\phi^nu^2=-3+\frac{3(2-\gamma)}{2}\Omega_m,\label{raycha-sbt'}\\
&&\frac{\ddot\phi}{\sqrt{6}H^2}=-3u-\sqrt\frac{3}{2}\frac{n}{\phi}u^2.\label{kg-sbt'}\eea Then, by taking the $\tau$-derivative of the variable $u$ one obtains,

\bea u'=-3u-\sqrt\frac{3}{2}\frac{n}{\phi}u^2-u\frac{\dot H}{H^2},\label{u'}\eea or, if take into account \eqref{raycha-sbt'},

\bea u'=-\sqrt\frac{3}{2}\frac{n}{\phi}u^2-\frac{3(2-\gamma)}{2}\Omega_m u,\label{u-ode}\eea where it is evident that this is not a closed equation (another variable is required). Next let us take the $\tau$-derivative of Eq. \eqref{fried-sbt'}:

\bea \Omega'_m=-2w\phi^n u\left(\sqrt\frac{3}{2}\frac{n}{\phi}u^2+u'\right),\label{om-der}\eea and substitute $u'$ from \eqref{u-ode} into \eqref{om-der}, we obtain

\bea \Omega'_m=3(2-\gamma)\omega\phi^n u^2\Omega_m.\label{qed}\eea Finally, if in this last equation take into account Eq. \eqref{fried-sbt'}: $\omega\phi^n u^2=1-\Omega_m$, then we obtain the same dynamical system \eqref{om-eq} of the EMS system. This means that both share the same asymptotic dynamics, which is what we wanted to demonstrate.

%%%%%%%%%%%%%%%%%%%%%%%%%%%%%%%%%%%%%%%%%%%%%%%%%%

\begin{table*}[tbh]\centering
\begin{tabular}{||c||c|c|c|c|c|}
\hline\hline
Critical Point  &  $\Omega_m$  & $\Omega_\Lambda$  & Existence & Stability & $q$\\
\hline\hline
$A$ (stiff-matter) & $0$ & $0$ & Always & Unstable (past attractor) & $2$ \\
\hline
$B$ (matter dominance) & $1$ & $0$ & '' & Saddle point & $-1+3\gamma/2$ \\
\hline
$C$ (de Sitter) & $0$ & $1$ & '' & Stable (future attractor) & $-1$ \\
\hline\hline\end{tabular}
\caption{Critical points of the ASODE \eqref{asode-l}, together with their existence and stability conditions, as well as the values of the deceleration parameter $q=-1-\dot H/H^2$.}\label{tab-1}\end{table*}

%%%%%%%%%%%%%%%%%%%%%%%%%%%%%%%%%%%%%%%%%%%%%%%%%%

%%%%%%%%%%%%%%%%%%%%%%%%%%%%%%%%%%%%%%%%%%%%%%%%%%%%%%%%%%%%%%%%%%%%%%%%%%%

\section{SBT and accelerated expansion}\label{sect-acc-expans}

%%%%%%%%%%%%%%%%%%%%%%%%%%%%%%%%%%%%%%%%%%%%%%%%%%%%%%%%%%%%%%%%%%%%%%%%%%%

The accelerated expansion of the universe can be explained within the SBT theory only if add a cosmological constant to the matter content as in GR theory \cite{roman-2014, indii-2022}. Actually, if add a cosmological constant term $\Lambda$ in the right-hand side of Eq. \eqref{fried-sbt} (the remaining equations are unchanged):

\bea 3H^2=\rho_m+\frac{\omega}{2}\phi^n\dot\phi^2+\Lambda,\label{cc-sbt}\eea and define the dimensionless energy density of the cosmological constant: $\Omega_\Lambda\equiv\Lambda/3H^2$, instead of equations \eqref{fried-sbt'} and \eqref{raycha-sbt'} one obtains that,

\bea &&\Omega_m+\Omega_\Lambda=1-\omega\phi^nu^2,\label{raycha-ol}\\
&&\frac{\dot H}{H^2}=-3(1-\Omega_\Lambda)+\frac{3(2-\gamma)}{2}\Omega_m,\label{om-ol}\eea respectively. If derive \eqref{raycha-ol} with respect to $\tau$ and take into account equations \eqref{u'} and \eqref{om-ol}, one obtains the following equation:

\bea \Omega'_m+\Omega'_\Lambda=6(1-\Omega_m-\Omega_\Lambda)\left(\frac{2-\gamma}{2}\Omega_m+\Omega_\Lambda\right).\label{om'-ol'}\eea Besides, since

\bea \Omega'_\Lambda=-2\Omega_\Lambda\frac{\dot H}{H^2},\nonumber\eea we finally obtain the following ASODE:

\bea &&\Omega'_m=3(2-\gamma)\Omega_m(1-\Omega_m)-6\Omega_m\Omega_\Lambda,\nonumber\\
&&\Omega'_\Lambda=6\Omega_\Lambda(1-\Omega_\Lambda)-3(2-\gamma)\Omega_\Lambda\Omega_m.\label{asode-l}\eea Notice that if set $\Omega_\Lambda=0$ we get the autonomous ODE \eqref{om-eq}. The phase space is defined as the following triangular region in the $(\Omega_m\Omega_\Lambda)$-plane:

\bea &&\Psi:\left\{(\Omega_m,\Omega_\Lambda)|\Omega_m-\Omega_\Lambda\leq 1,\right.\nonumber\\
&&\left.\;\;\;\;\;\;\;\;\;\;\;\;\;\;\;\;\;\;\;\;\;\;\;\;0\leq\Omega_m\leq 1,\,0\leq\Omega_\Lambda\leq 1\right\}.\label{ph-space}\eea

%-----------------------------------

\begin{figure}[t!]
\includegraphics[width=6cm]{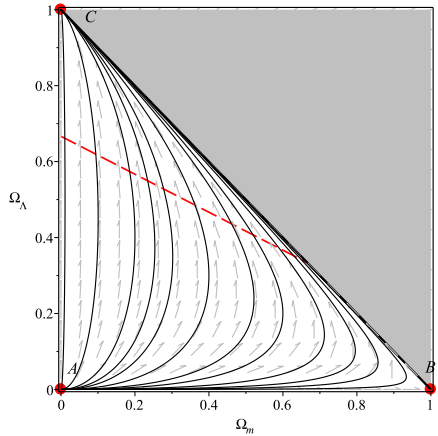}
\vspace{1.2cm}\caption{Phase portrait of the dynamical system \eqref{asode-l}. Here we consider dust matter with $\gamma=1$. The critical points $A:(0,0)$ (stiff-matter solution), $B:(1,0)$ (matter dominance) and $C:(0,1)$ (de Sitter solution), are enclosed by the small solid circles. The dashed curve corresponds to the condition ($q=-1-\dot H/H^2=0$). Above the dashed curve the expansion is accelerated while below it the cosmic expansion is decelerated.}\label{fig1}\end{figure}

%----------------------------------------

The critical points of the dynamical system \eqref{asode-l} are shown in TAB. \ref{tab-1}, where the stability properties and the values of other physical parameters are shown as well. The corresponding phase portrait for the case when the background matter is dust ($\gamma=1$) is shown in FIG. \ref{fig1}. The critical points $A$ (stiff-matter solution), $B$ (matter-dominated solution) and $C$ (de Sitter solution) are enclosed by the small circles. The dash (tilted) line represents the boundary where $q=-1-\dot H/H^2=0$. Above this line the expansion is accelerated, while below it it is decelerated instead. It is seen that every orbit in the phase space $\Psi$ starts at the past attractor (point $A$) and ends up at the future de Sitter attractor (point $C$). 

The stiff-matter solution $A$ is characterized by:

\bea 3H^2=\frac{\omega}{2}\phi^n\dot\phi^2,\;\frac{\dot H}{H^2}=-3,\nonumber\eea from where it follows that

\bea \phi(t)=\left[\ln\left(t-t_0\right)^\frac{n+2}{\sqrt{6\omega}}\right]^\frac{1}{n+2},\label{stiff-sol}\eea where $t_0$ is an integration constant. Meanwhile, the matter-dominated phase $B$ is a transient stage of the cosmic expansion since $B$ is a saddle critical point. In this case,

\bea \frac{\dot H}{H^2}=-\frac{3\gamma}{2}\;\Rightarrow\;a(t)=a_0(t-t_0)^{2/3\gamma},\label{matter-sol}\eea where $a_0$ is another integration constant. Finally, the de Sitter solution $C$ describes an accelerated expansion stage where $a(t)=a_0\exp{[H_0(t-t_0)]}$.

%----------------------------------------

But for the stiff-matter critical point, the asymptotic behavior of a GRT-based cosmological model with the cosmological constant is the same as the one we have obtained for the SBT-based model, with the difference that GRT is a more simple framework (one less field degree of freedom.) Hence, although accelerated expansion is an asymptotic stable state of the dynamical system \eqref{asode-l}, this behavior has been obtained at the cost of losing the original motivation for replacing the much more simple GRT by the SBT based cosmological model.

%-------------------------------------

%%%%%%%%%%%%%%%%%%%%%%%%%%%%%%%%%%%%%%%%%%%%%%%%%%

\begin{table*}[tbh]\centering
\begin{tabular}{||c||c|c|c|c|c|c|}
\hline\hline
Critical Point  &  $\Omega_m$  & $\Omega_\Lambda$  & Existence & Stability & $q$ & $\Omega_m$\\
\hline\hline
$A_+$ (stiff-matter) & $1$ & $0$ & Always & Unstable (past attractor) if $\lambda<1$ & $2$ & $0$ \\
&&&& Saddle if $\lambda>1$ && \\
\hline
$A_-$ (stiff-matter) & $-1$ & $0$ & '' & Unstable (past attractor) & $2$ & $0$ \\
\hline
$B$ (matter dominance) & $0$ & $0$ & '' & Saddle point & $-1+3\gamma/2$ & $1$ \\
\hline
$D$ ($\phi$-dominance) & $\lambda$ & $\sqrt{1-\lambda^2}$ & $|\lambda|<1$ & Stable (future attractor) if $\gamma>2\lambda^2$ & $-1$ & $0$ \\
&&&& Saddle if $\gamma<2\lambda^2$ && \\
\hline
$S$ (matter-scaling) & $\frac{\gamma}{2\lambda}$ & $\frac{\sqrt{\gamma(2-\gamma)}}{2\lambda}$ & $\lambda>\sqrt\frac{\gamma}{2}$ & Stable if $f_1<1$ and $f_2>0$ & $-1$ & $1-\frac{\gamma}{2\lambda^2}$ \\
&&&& Stable spiral if $f_1<1$ and $f_2<0$ $\lambda>1$ && \\
&&&& Saddle if $f_1>1$ && \\
\hline\hline\end{tabular}
\caption{Critical points of the ASODE \eqref{asode-v}, together with their existence and stability conditions, as well as the values of the deceleration parameter ($q=-1-\dot H/H^2$) and of the dimensionless energy density of matter ($\Omega_m$). In the last row we introduced the following shorthand notation: $f_1=f_2/(2-\gamma)$, $f_2\equiv 4\gamma^2-9\lambda^2\gamma+2\lambda^2$.}\label{tab-2}\end{table*}

%%%%%%%%%%%%%%%%%%%%%%%%%%%%%%%%%%%%%%%%%%%%%%%%%%

%-----------------------------------

\begin{figure*}[t!]
\includegraphics[width=5cm]{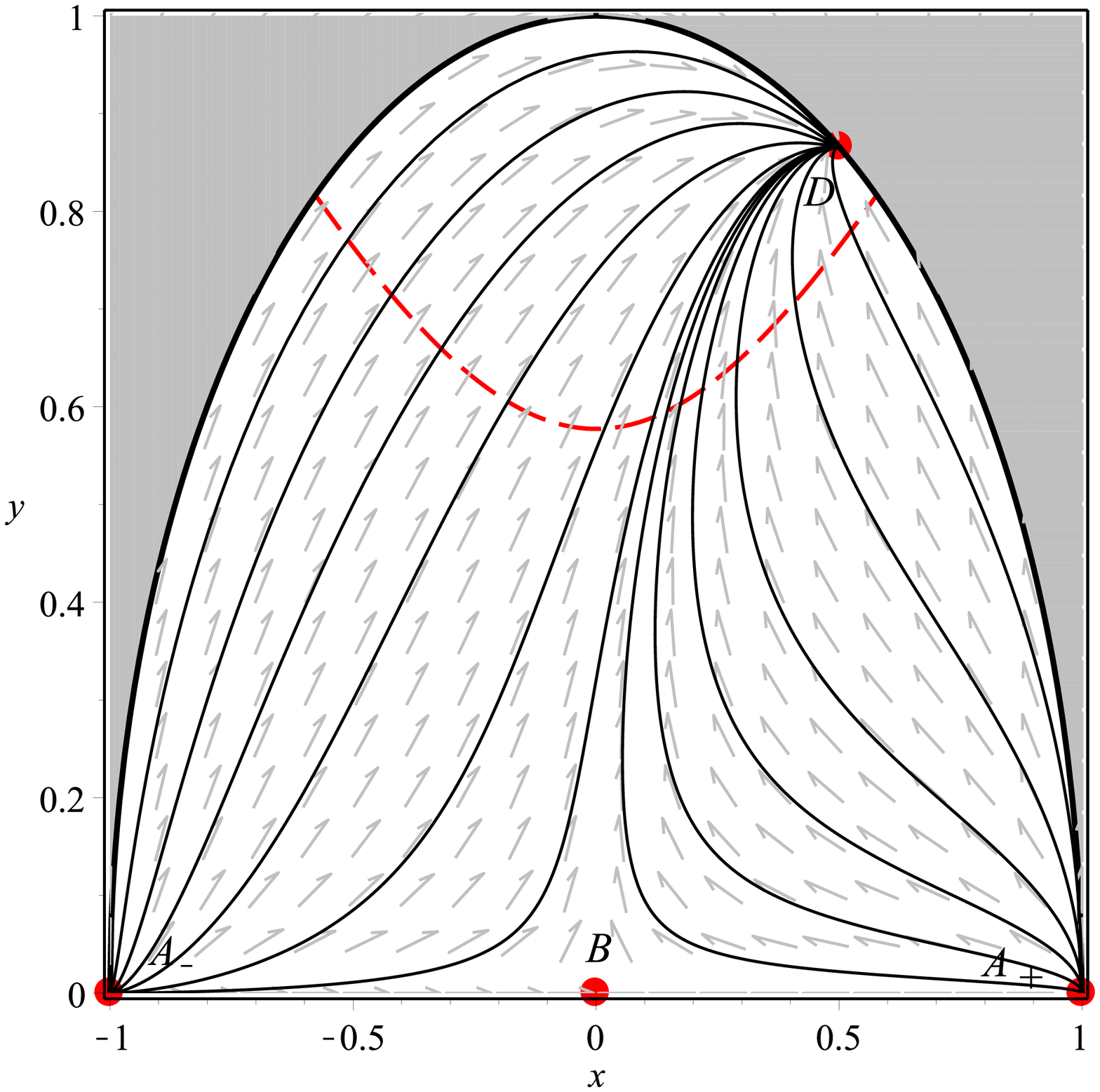}
\includegraphics[width=5cm]{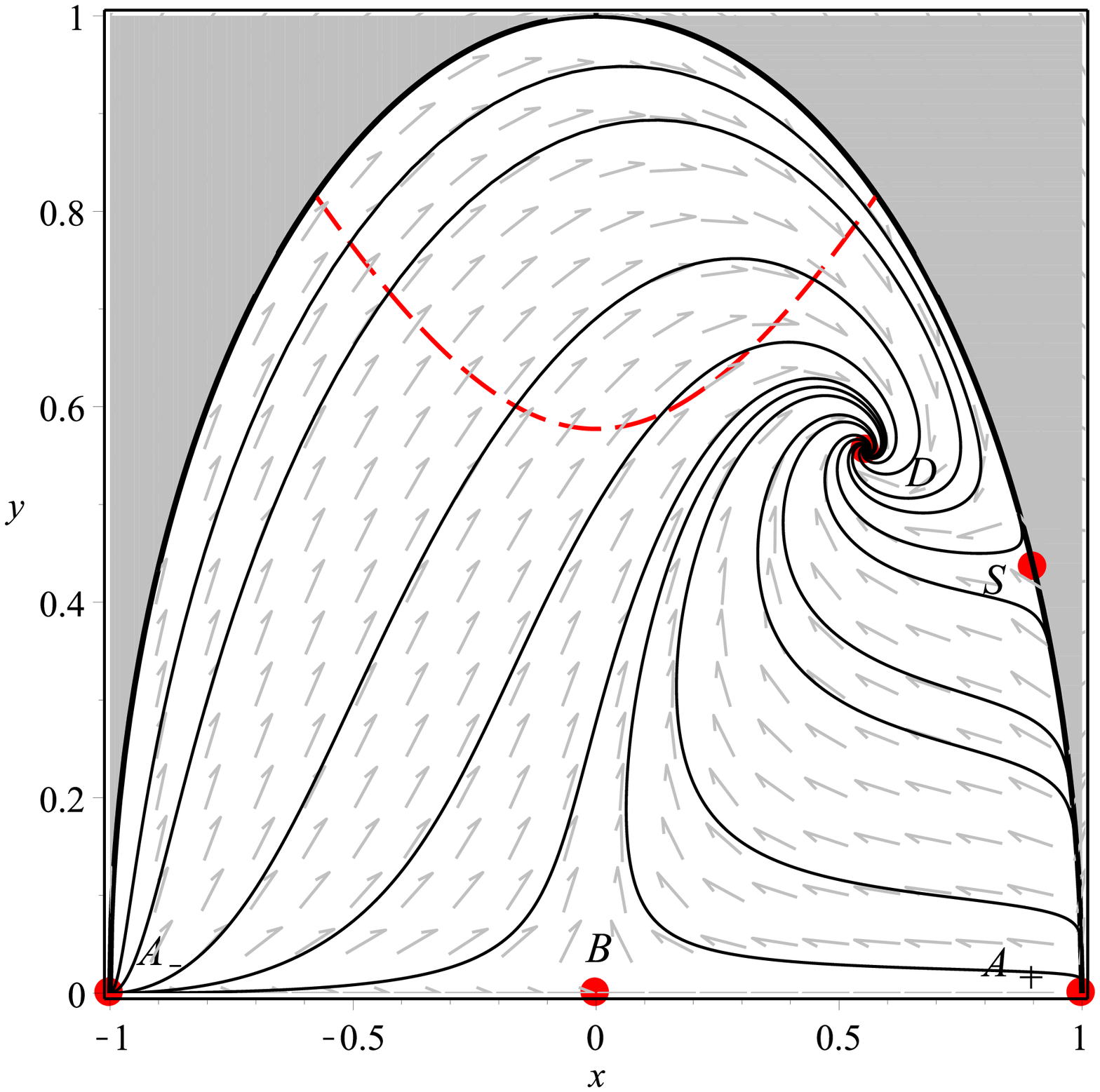}
\includegraphics[width=5cm]{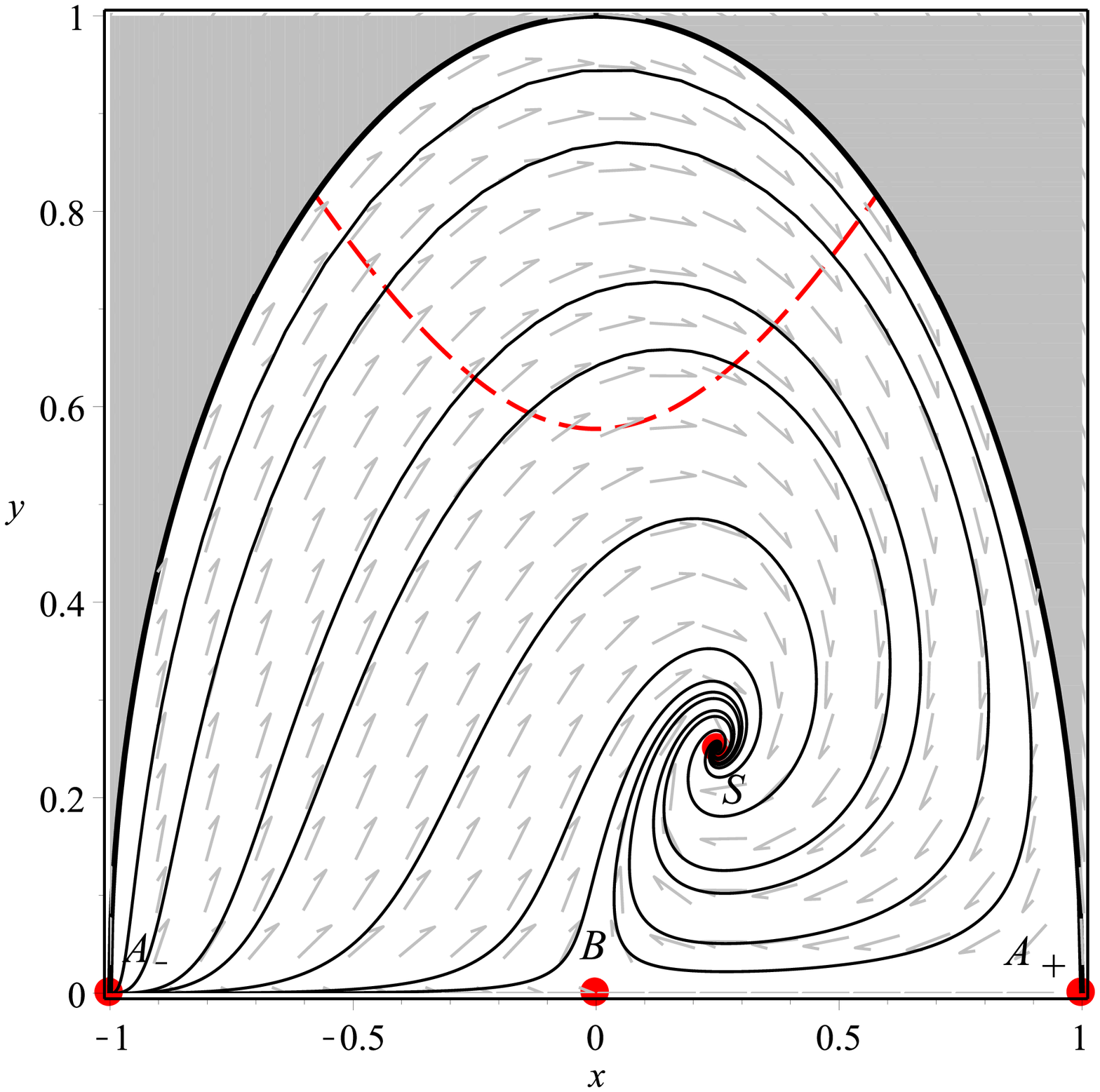}
\vspace{1.2cm}\caption{Phase portrait of the dynamical system \eqref{asode-v}. Here we consider dust matter with $\gamma=1$. The left figure is for $\lambda=0.5$, while the middle figure is for $\lambda=0.9$ and the right figure is for $\lambda=2$. The critical points $A_\pm:(\pm 1,0)$ (stiff-matter solutions), $B:(0,0)$ (matter dominance), $D:(\lambda,\sqrt{1-\lambda^2})$ (scalar field dominance) and $S:(\gamma/2\lambda,\sqrt{\gamma(2-\gamma)}/2\lambda)$ (matter-scaling solution), are enclosed by the small solid circles. The dashed curve corresponds to the condition $q=-1-\dot H/H^2=0$. Above the dashed curve the expansion is accelerated while below it the cosmic expansion is decelerated.}\label{fig2}\end{figure*}

%----------------------------------------

%%%%%%%%%%%%%%%%%%%%%%%%%%%%%%%%%%%%%%%%%%%%%%%%%%%%%%%%%%%%%%%%%%%%%%%%%%%%%%%%

\section{Self-interacting scalar field in the SBT theory}\label{sect-self-sbt}

%%%%%%%%%%%%%%%%%%%%%%%%%%%%%%%%%%%%%%%%%%%%%%%%%%%%%%%%%%%%%%%%%%%%%%%%%%%%%%%%%

For completeness of the present study, in this section we include a dynamical systems investigation of the SBT-based cosmological model when the scalar field is a self-interacting one. I. e., in the gravitational action \eqref{sbt-action} a self-interaction potential $V=V(\phi)$ for the scalar field is added:
 
\bea S_g=\int d^4x\sqrt{-g}\left[R-\omega\phi^n(\der\phi)^2-2V\right].\label{sbt-v-action}\eea The following EOM are obtained from this action (compare with \eqref{sbt-eom}):

\bea &&G_{\mu\nu}=\omega\phi^n\left[\der_\mu\phi\der_\nu\phi-\frac{1}{2}g_{\mu\nu}(\der\phi)^2\right]-g_{\mu\nu}V,\nonumber\\
&&\nabla^2\phi+\frac{n}{2}\frac{(\der\phi)^2}{\phi}=\frac{V_{,\phi}}{\omega\phi^n},\label{sbt-v-eom}\eea where we have introduced the shorthand notation $V_{,\phi}\equiv dV/d\phi$. In terms of the FRW metric \eqref{frw-metric} these equations read (compare with the cosmological equations \eqref{fried-sbt}-\eqref{kg-sbt}):

\bea &&3H^2=\rho_m+\frac{\omega}{2}\phi^n\dot\phi^2+V,\label{fried-sbt-v}\\
&&2\dot H=-\gamma\rho_m-\omega\phi^n\dot\phi^2,\label{raycha-sbt-v}\\
&&\ddot\phi+3H\dot\phi+\frac{n}{2}\frac{\dot\phi^2}{\phi}=-\frac{V_{,\phi}}{\omega\phi^n},\label{kg-sbt-v}\eea where a matter fluid has been considered. The corresponding continuity equation reads $\dot\rho_m+3\gamma H\rho_m=0$. If introduce the following variables of some states space:

\bea x\equiv\sqrt\frac{\omega}{6}\phi^\frac{n}{2}\frac{\dot\phi}{H},\;y\equiv\frac{\sqrt{V}}{\sqrt{3}H},\label{v-vars}\eea the above EOM-s can be traded by the following ASODE (as before the tilde means derivative with respect to the number of e-foldings $\tau=\ln a(t)$):

\bea &&x'=-3x\left[1-x^2-\frac{\gamma}{2}\left(1-x^2-y^2\right)\right]-\frac{3V_{,\phi}/V}{\sqrt{6\omega}\phi^{n/2}}\,y^2,\nonumber\\
&&y'=3y\left[\frac{V_{,\phi}}{V}\frac{\dot\phi}{6H}+x^2+\frac{\gamma}{2}\left(1-x^2-y^2\right)\right],\label{asode-v-1}\eea where we assume expanding universe exclusively, so that $y\geq 0$, and the Friedmann constraint,

\bea \Omega_m=1-x^2-y^2,\label{fried-v}\eea as well as the following equation:

\bea \frac{\dot H}{H^2}=-3x^2-\frac{3\gamma}{2}\left(1-x^2-y^2\right),\label{hdot-v}\eea have been taken into account. In general \eqref{asode-v-1} are not a closed system of ODE, so that additional phase space variables are required. Nevertheless, for constant self-interaction potential $V=V_0$ $\Rightarrow V_{,\phi}=0$, and for potentials of the following type:

\bea &&\frac{V_{,\phi}}{V}=-\sqrt{6\omega}\lambda\phi^\frac{n}{2}\;\Rightarrow\nonumber\\
&&V(\phi)=V_0\exp{\left(-\frac{2\sqrt{6\omega}\lambda}{n+2}\,\phi^{1+n/2}\right)},\label{v-pot}\eea where $\lambda$ is a non-negative constant, no more variables than $x$ and $y$ are necessary and the obtained ASODE:

\bea &&x'=-3x\left[1-x^2-\frac{\gamma}{2}\left(1-x^2-y^2\right)\right]+3\lambda y^2,\nonumber\\
&&y'=3y\left[-\lambda x+x^2+\frac{\gamma}{2}\left(1-x^2-y^2\right)\right],\label{asode-v}\eea is a closed system of ODE. Here we shall investigate the cosmological dynamics of the SBT-based cosmological model with self-interaction potential \eqref{v-pot}.

The critical points of the dynamical system \eqref{asode-v}, which are found in the physical phase space

\bea \Psi:\{(x,y)|x^2+y^2\leq 1,\;|x|<1,\;0\leq y\leq 1\},\nonumber\eea together with their existence and stability properties, as well as the values of the deceleration parameter $q$ and of the dimensionless energy density of matter $\Omega_m$ are summrized in TAB. \ref{tab-2}. Meanwhile, in FIG. \ref{fig2} the phase portrait corresponding to this ASODE is drawn for dust ($\gamma=1$), for three different values of the parameter $\lambda$: $\lambda=0.5$ (left figure), $\lambda=0.9$ (middle figure) and $\lambda=2$ (right figure).

At first sight it might seem noticeable that the obtained asymptotic dynamics in this case, is the same as the one for the Einstein-self-interacting-scalar model \eqref{fried-eq-ems}--\eqref{cont-eq} with the addition of an exponential self-interaction potential $V(\vphi)=V_0\exp{(-\lambda\vphi)}$, which has been discussed in detail in Ref. \cite{wands-prd-1998}. However, a closer look at Eq. \eqref{sf-redef} shows us that the, if we make this scalar field redefinition in Eq. \eqref{v-pot}, what one obtains is, precisely, the exponential potential for the new scalar field:

\bea V(\vphi)=V_0\exp{(-\sqrt{6}\lambda\vphi)},\nonumber\eea where we need to make the replacement $\lambda\rightarrow\sqrt{6}\lambda$ in order to exactly meet the results of \cite{wands-prd-1998}. This demonstrates that the dynamical equivalence between the SBT and the EMS goes beyond massless scalar fields and also holds true in the presence of self-interactions.

%--------------Discussion-----------------

%%%%%%%%%%%%%%%%%%%%%%%%%%%%%%%%%%%%%%%%%%%%%%%%%%%%%%%

\section{Discussion and conclusion}\label{sect-discu}

%%%%%%%%%%%%%%%%%%%%%%%%%%%%%%%%%%%%%%%%%%%%%%%%%%%%%%

Although we have already mentioned that the S\'aez-Ballester theory is not a scalar-tensor theory of gravity, let us further discuss on this subject. First we need to answer the following question: what is a STT of gravity? As suggested by its name, in a scalar-tensor theory of gravity both the metric and the scalar field are propagators of the gravitational interactions. This is reflected, in particular, in the measured value of the Newton's constant $G_N$. In the introduction we have shown this in the case of Brans-Dicke (BD) type scalar-tensor theories of gravity. Below we go a step further and we shall show what a STT is in a most generalized case.

Among the most general scalar-tensor theories of gravity are those which are included in the Horndeski classification \cite{quiros-ijmpd-rev-2019, clifton-phys-rept-2012, quiros-ijmpd-2020, horn-1974, charm-prl-2012, kobayashi-rev-2019}. The Horndeski class is given by the following action,

\bea &&S_H=\int d^4x\sqrt{-g}\left[G_4R+K-G_3(\nabla^2\phi)\right.\nonumber\\
&&\;\;\;\;\;\;\;\;\;\;\;\;\;\;\;\;\;\;\;\;\;\;\;\;\;\;\;\;\;\;\;\left.+G_5G_{\mu\nu}\nabla^\mu\nabla^\nu\phi\right],\label{horn-action}\eea where $G_{\mu\nu}=R_{\mu\nu}-g_{\mu\nu}R/2$ is the Einstein's tensor, $K=K(\phi,X)$ and $G_3=G_3(\phi,X)$ are functions of the scalar field and of its kinetic energy density, while, for simplicity, here we assume that $G_4=G_4(\phi)$ and $G_5=G_5(\phi)$ can be functions of the scalar field exclusively. Although the Horndeski class \eqref{horn-action} includes self-interacting Einstein-scalar system (recall that in our units system $8\pi G_N=1$,)

\bea K=X-V(\phi),\;G_3=G_5=0,\;G_4=\frac{1}{2},\label{sies}\eea and the k-essence theories,

\bea K=f(\phi)g(X),\;G_3=G_5=0,\;G_4=\frac{1}{2},\label{kess}\eea which are not STTs, scalar-tensor theories beyond BD-type are also included. For instance \cite{quiros-ijmpd-rev-2019, quiros-ijmpd-2020}:

\begin{itemize}

\item{\it BD theory.}

\bea K=\frac{\omega_{BD}}{\phi}\,X-V(\phi),\;G_3=G_5=0,\;G_4=\phi,\nonumber\eea where $\omega_{BD}$ is the BD coupling constant.

\item{\it Cubic galileon in the Einstein frame.}

\bea G_3=2\sigma X,\;G_4=\frac{1}{2},\;G_5=0,\nonumber\eea where $\sigma$ is the cubic self-coupling.

\item{\it Kinetic coupling to the Einstein's tensor.}

\bea K=X-V,\;G_3=0,\;G_4=\frac{1}{2},\;G_5=-\frac{\alpha}{2}\,\phi,\nonumber\eea where $\alpha$ is the coupling constant.

\end{itemize} 

In order to demonstrate that the above are STTs let us write the expression for the measured (Cavendish-type) gravitational constant. According to \cite{hohmann-prd-2015} for those Horndeski theories where the PPN formalism can be applied we get that,

\bea 8\pi G_\text{cav}=\frac{1}{2G_4}\left[\frac{3G^2_{4,\phi}+G_4K_{,X}+G^2_{4,\phi}\,e^{-Mr}}{3G^2_{4,\phi}+G_4K_{,X}}\right],\label{g-cav}\eea where $Y_{,X}\equiv dY/dX$, $Z_{,\phi\phi}\equiv d^2Z/d\phi^2$, etc., and 

\bea M=\sqrt{\frac{-2G_4K_{,\phi\phi}}{K_{,X}+3G^2_{4,\phi}}}.\nonumber\eea Above we have taken into account that for the cubic galileon a Vainshtein-like screening takes place \cite{dearcia-2018} so that the PPN formalism can not be applied in this case. For that reason we have set $G_3=0$. For kinetic coupling theory \eqref{g-cav} is not valid either, although in this case the coupling of the derivative of the scalar field to the curvature through $G_{\mu\nu}$ already suggests that it is a STT. For further explanation why the cubic galileon and the kinetic coupling theory are actually scalar-tensor theories of gravity we recommend the discussion in \cite{quiros-ijmpd-rev-2019}. 

From equation \eqref{g-cav} it is evident that for constant $G_4=1/2$, the measured gravitational constant coincides with the Newton's constant $G_\text{cav}=(8\pi)^{-1}$, as in GR. This result holds true for any $K=K(\phi,X)$, so that k-essence may be identified with general relativity with an exotic scalar field as source of the Einstein's equations. This includes, of course, the EMS system.

Another subject which is to be taken into account when one discuss about equivalence of gravitational theories with the inclusion of scalar fields and the classification of the STT, is the one concerning the conformal transformations of the metric. However, this discussion goes beyond the purpose and the reach of the present paper and we postpone it for a forthcoming publication. 

We expect that the present discussion will suffice to amend a frequent and long standing misconception. As a matter of fact S\'aez-Ballester theory is not an independent theory since it the same as EMS theory. SBT was proposed in a 1986 paper \cite{s-b-theory} while, as long as we know, EMS system was introduced as early as in 1957 year \cite{bergmann-prd-1957}. By the time when \cite{s-b-theory} was published dozen of papers on EMS already existed in scientific bibliography. What is worse is to incorrectly classify the SBT as a scalar-tensor theory. We hope this misunderstanding would be fixed as well.

%--------------------------------------------------------

%------------acknowledgments-----------------

{\bf Acknowledgments.} We thank Mar\'ia Jos\'e Guzm\'an and Andronikos Paliathanasis for encouraging comments and Kazuharu Bamba and Vasilis Oikonomou for pointing to us bibliographic references. We also acknowledge FORDECYT-PRONACES-CONACYT for support of the present research under grant CF-MG-2558591.

%%%%%%%%%%%%%%%%%%%%%%%%%%%

%%%%%%%%%%%%%%%%%%%%%%%

%%%%%%%%%%%%%%%%
%%%%%%%%%%%%%%%%

\end{document}